\documentclass
[showpacs,preprintnumbers,prb,twocolumn,byrevtex,letterpaper,groupedaddress,10pt]{revtex4}%
\usepackage{amsmath}
\usepackage{amssymb}
\usepackage{graphicx}
\usepackage{dcolumn}
\usepackage{bm}
\usepackage{acronym}
\usepackage{amsfonts}%
\providecommand{\U}[1]{\protect\rule{.1in}{.1in}}
\providecommand{\U}[1]{\protect\rule{.1in}{.1in}}
\begin{document}
\preprint{RYKOV Alexandre / \ February 4 (2013)}
\title{MIXED ORBITAL GROUND STATES OF Fe$^{2+}$ IN PRUSSIAN BLUES}
\author{A. I. Rykov,$^{1}$ J. Wang$^{1}$, T. Zhang$^{1}$, and K. Nomura$^{2}$ }
\affiliation{$^{1}$M\"{o}ssbauer Effect Data Center, Dalian Institute of Chemical Physics,
Chinese Academy of Sciences, 457 Zhongshan Road, Dalian 116023, China, $^{2}%
$School of Engineering, The University of Tokyo, Hongo 7-3-1, 113-8656, Japan,}
\date{\today}

\begin{abstract}
We report on the mixed-orbital ground state compounds analogous to the
mixed-valence ones. Orbital doublet and singlet ground states of the Fe$^{2+}$
ion displayed in M\"{o}ssbauer spectra of the Prussian blue analogues A$_{y}%
$Fe$_{3-y}$[Co(CN)$_{6}$]$_{2}$ \textperiodcentered$x$H$_{2}$O ($y\leqslant1$)
are interconverting to each other as temperature changes for A = K, Na.
Relative weight of orbital singlet ground states increases with lowering
temperature. In the alcali-free cobalticyanides, the M\"{o}ssbauer spectra are
dominated by quadrupole-splt doublets of the same origin with a large
splitting, characteristic of the Fe$^{2+}$ species coordinated by two (at
least) oxygen atoms of water molecules in \textit{cis}-configuration.
Single-type octahedral Fe$^{2+}$ coordination is isolated in A-filled
cobalticyanides for A=Rb, Cs to be characterized by the narrower quadrupole
spectrum associated with the orbital doublet ground states. Ionic exchange of
K for Cs in KFe[Co(CN)$_{6}$] \textperiodcentered H$_{2}$O results in
predominance of the orbital singlet, i.e., in enhanced anisotropy of the
Fe$^{2+}$ valence electrons. We found hence that the Fe$^{2+}$charge
distribution can be modified together with crystal size at the step of
synthesis owing to distortion isomerism.

\end{abstract}

\maketitle

\makeatletter\global\@specialpagefalse\let\@evenhead\@oddhead
\let\@evenfoot\@oddfoot\makeatother\label{sec:intro}

\section{Introduction}

Materials enclosing the transition metal elements are characterized by a wide
diversity of ground states whether this be the quantum states of collective
electron systems or local states. In latter case, the transitions between
disparate ground states of 3d ions can be driven by changes in temperature,
chemical doping, or by external stimuli. Phase transitions are thus induced
that involve the spin, charge, lattice and orbital degrees of freedom. Charge
transfer, spin crossover and linkage isomerism are the three types of
transitions requiring a complex elucidation of the roles of corresponding
degrees of freedom to be distinguished from each
other\cite{Kato,GutlichCSR,LI}. The long-range structural aspects display
their role via succession of transition steps as they evidence the
interconversions between various local states mediated by elastic
interactions\cite{Chong,Koudr}. Charge transfer from one site to another could
be a prerequisite of the concomitant spin-state
transition\cite{Moritomo,Kabir}. Orbital degree of freedom is responsible for
a long-range-distortion symmetry lowering, provided that orbital ordering is
coupled with the mixed-valence charge ordering transition\cite{Rykov}. Orbital
ground state reversal
\cite{Dezsi,Coey,Reiff1973,Latorre,Latorre3,Latorre2,Volland2,Volland,Brunot,Isotope}
is an example of yet another type of transitions strongly coupled to the
lattice and, possibly, to some other degrees of freedom. Simultaneous presence
of the orbital singlet and orbital doublet ground states, familiar in case of
Fe$^{2+}$ ion\cite{Hoy}, could be an indication of proximity of such a transition.

Prussian blue (PB) analogues constitute a family of simple cubic compounds
which excited a great interest recently owing to the family members
exemplifying the spin crossover, charge transfer and linkage isomerism
\cite{Kato,GutlichCSR,LI}. Here, in the same family, A$_{y}$Fe$_{3-y}%
$[Co(CN)$_{6}$]$_{2}$ \textperiodcentered$n$H$_{2}$O ($y\leqslant1$, A = K,
Na, Rb, Cs), we report on\ mixing and intercoverting the orbital ground states
of Fe$^{2+}$.

Metal-organic framework of the PB analogues is formed by the 3d metal cations
and the cyanometallate complex anions (of the [Co(CN)$_{6}$]$^{3-}$-type)
arranged into a NaCl-like 3D checkerboard. In major group of these compounds,
the valence and spin differ between the states of metal ions occupying the
cationic node and the anion core. The framework complex anions are typically
having higher valences than the cations, in this case, the framework is
bearing a net negative charge, which is balanced either by insertion of alcali
(A) cations into the framework interstitials or by spontaneous generation of
the large-size vacancies on the anionic nodes. Crystallization and zeolitic
water molecules (C-water and Z-water) fill the cavities at the vacant sites of
complex anions and A-cations, respectively.

The local electronic configurations are probed by the quadrupole interactions
of a $^{57}$Fe M\"{o}ssbauer nuclei sensing the electric field gradient from
valent electrons and lattice charges surrounding the Fe$^{2+}$ ion.
M\"{o}ssbauer spectra of ferrous hexacyanocobaltates were reported previously
only for the alcali-free phases ($y=0$)\cite{Ras},\cite{Reg}. The spectra can
be fitted by two or more quadrupole doublets implying the existence of a
number of different states of the high-spin Fe$^{2+}$ ions. Paired doublets,
both originating from the cationic sites outside the low-spin complex
[M(CN)$_{6}$] were also observed in M\"{o}ssbauer spectra of some other
ferrous hexacyanometallates, e.g., for M=Cr ($y=0$)\cite{survey}. The early
M\"{o}ssbauer study\cite{Ras} has proposed that the water molecules in
\textit{cis-} and \textit{trans-} configurations could contribute to the
appearance of multiple sites for Fe$^{2+}.$ In case of M=Cr, the spin-state
transition of Fe$^{2+}$was observed only for $y\neq0$ (A=Cs) \cite{Nomura}. It
was conjectured\cite{Nomura} that filling the structure with A ions induces a
stronger ligand field on the Fe$^{2+}$ ion that causes the spin transition. In
this study of the local electronic configurations in AFe[Co(CN)$_{6}$]$\cdot
$H$_{2}$O ( A = K, Na, Rb, Cs) and related alcali-free form we show that the
ligand field stabilizes the singular ground state (orbital doublet) only for
A=Cs and Rb. In the cases of A=Na, K, and $y=0$ we report on the coexisting
orbital singlet and orbital doublet ground states for Fe$^{2+}$ and on the
interconversions between them induced by changes of temperature.

\section{ Experimental}

RbFe[Co(CN)$_{6}$]$\cdot$H$_{2}$O and CsFe[Co(CN)$_{6}$]$\cdot$H$_{2}$O were
prepared mixing the 0.025 M aqueous solution of K$_{3}$Co(CN)$_{6}$with 0.025
M solutions of FeCl$_{2}\cdot$4H$_{2}$O in excess of RbCl or CsCl. Similarly
synthesized samples in excess of KI and NaCl gave the compounds K$_{0.8}%
$Fe$_{1.1}$[Co(CN)$_{6}$]$\cdot2$H$_{2}$O and Na$_{0.7}$Fe$_{1.15}%
$[Co(CN)$_{6}$]$\cdot3$H$_{2}$O, respectively. The A-free compounds were
synthesized from solutions of similar molarity (sample 8 in Table I) and from
10-fold diluted (sample 6). Analytic methods of inductively-coupled plasma and
x-ray-fluorescent analysis gave the results consistent with the site
occupancies obtained from Rietveld analysis. The same compounds with different
crystal size were synthesized for A=Rb and Cs via the ionic exchange using
KFe[Co(CN)$_{6}$]$\cdot2$H$_{2}$O as a sorbent. Transformation of this sorbent
into CsFe[Co(CN)$_{6}$]$\cdot$H$_{2}$O was detected by Rietveld analysis and
confirmed by analytic techniques.

\textbf{Table I.} Lattice parameters in the precipitated and ion-exchanged
hexacyanocobaltates at room temperature.%

\begin{tabular}
[c]{l}%
\\\hline
\multicolumn{1}{|l|}{1}\\\hline
\multicolumn{1}{|l|}{2}\\\hline
\multicolumn{1}{|l|}{3}\\\hline
\multicolumn{1}{|l|}{4}\\\hline
\multicolumn{1}{|l|}{5}\\\hline
\multicolumn{1}{|l|}{6}\\\hline
\multicolumn{1}{|l|}{7}\\\hline
\multicolumn{1}{|l|}{8}\\\hline
\end{tabular}%
\begin{tabular}
[c]{|l|l|}\hline
Compound & Parameter*, nm\\\hline
Na$_{0.7}$Fe$_{1.15}$[Co(CN)$_{6}$]$\cdot3$H$_{2}$O &
\multicolumn{1}{|c|}{1.03278(8)}\\
K$_{0.8}$Fe$_{1.1}$[Co(CN)$_{6}$]$\cdot2$H$_{2}$O &
\multicolumn{1}{|c|}{1.03115(3)}\\
KFe[Co(CN)$_{6}$]$\cdot$H$_{2}$O & \multicolumn{1}{|c|}{1.03107(3)}\\
RbFe[Co(CN)$_{6}$]$\cdot$H$_{2}$O & \multicolumn{1}{|c|}{1.03455(7)}\\
CsFe[Co(CN)$_{6}$]$\cdot$H$_{2}$O & \multicolumn{1}{|c|}{1.03682(8)}\\
Fe[Co(CN)$_{6}$]$_{0.67}\cdot5$H$_{2}$O & \multicolumn{1}{|c|}{1.02778(8)}\\
Cs-exchanged (b) at 20$^{o}$C & \multicolumn{1}{|c|}{1.03628(8)}\\
$^{57}$Fe[Co(CN)$_{6}$]$_{0.67}\cdot5$H$_{2}$O &
\multicolumn{1}{|c|}{1.02813(9)}\\\hline
\end{tabular}

*Refined according to Fig.1 (c) and (d) structure models using the
Fm$\overline{3}$m space group.

X-ray Diffraction (XRD) patterns were measured on a PW3040/60 X'pert PRO
(PANalytical) diffractometer using Cu Ka source. The atomic ratio Fe/Co and
K/Co were determined by inductively coupled plasma atomic emission
spectrometer (ICP-AES) on an IRIS Interpid II XSP Instrument (Thermo Electron
Corporation). The x-ray fluorescent analysis was carried out using the XRF
Axios X (PANalytical) spectrometer. The water content in the samples was
estimated by thermogravimetric analysis carried out using a Setaram Setsys
16/18 analyser measuring the weight loss in an air flow of 25 ml/min with a
heating rate of 10 K/min. Magnetization of the paramagnetic compounds
was\ measured with a SQUID magnetometer in the range between 2 K to 350K.

M\"{o}ssbauer spectra were measured with a Topologic 500 A spectrometer at the
temperatures between 78 K and room temperature. The measurements at heating
the sample above room temperature were performed in a Wissel GMBH furnace.
Isomer shifts (IS) are reported relative to $\alpha$-Fe at room temperature.%

\begin{figure}
[ptb]
\begin{center}
\includegraphics[
height=4.5913in,
width=3.4056in
]%
{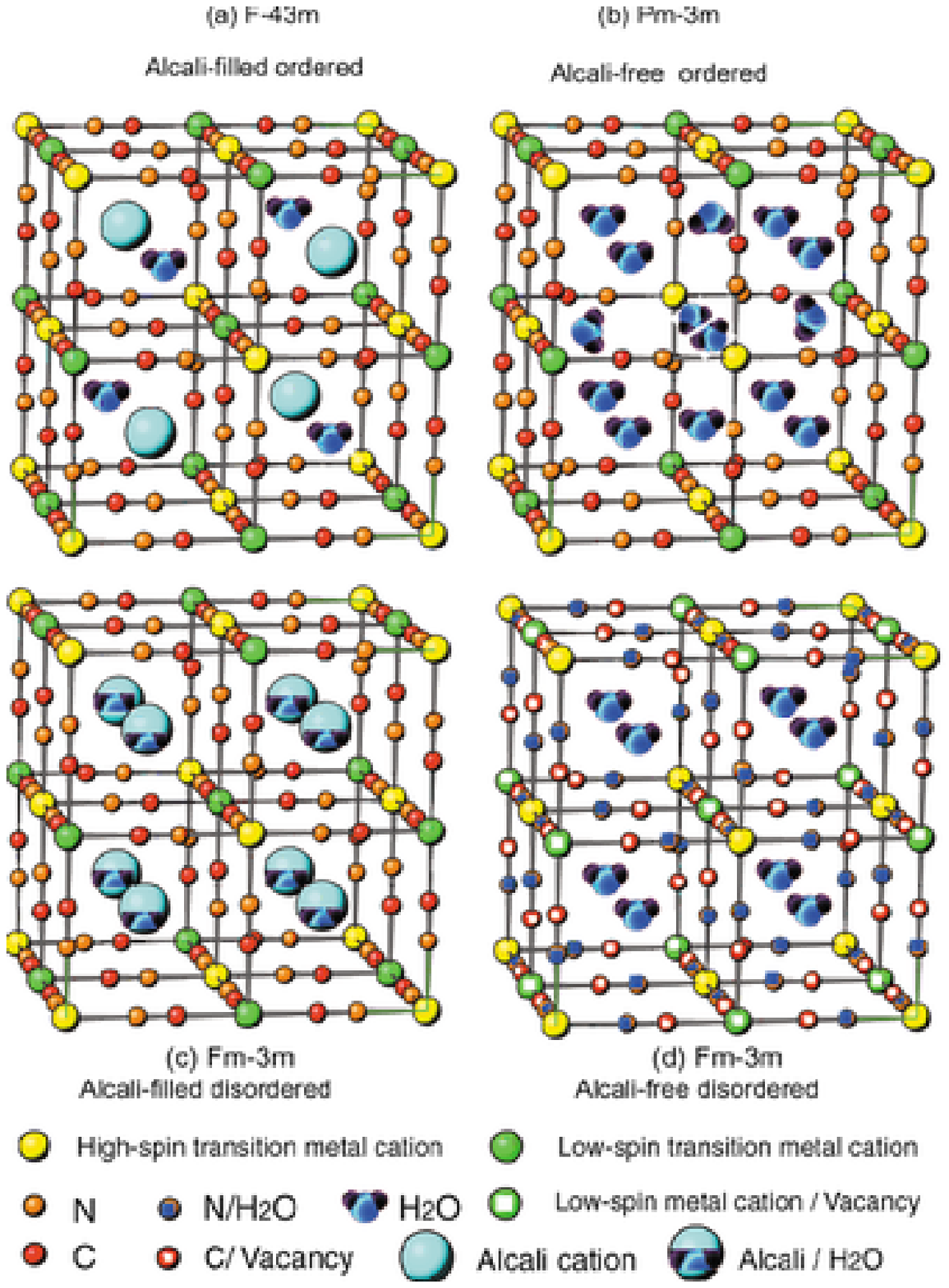}%
\caption{Crystal structures and symmetry groups of the four modifications of
Prussian blue analogues: alcali-filled ordered F$\overline{4}$3m (a),
alcali-free ordered Pm$\overline{3}$m (b), alcali filled disordered
Fm$\overline{3}$m (c) and alcali-free disordered Fm$\overline{3}$m (d). }%
\end{center}
\end{figure}

\section{Results and discussion}

\subsection{Structural considerations}

The A-filled and A-free Prussian Blues are two types of the cubic frameworks,
described by structure models of Keggin and Miles\cite{Keggin} and
Ludi\cite{Ludi}, respectively. Both of them exhibit the variable degrees of
randomness in the distribution of the A ions over the framework interstitials
and in the distribution of the water-filled anionic vacancies. In both cases,
the \textit{random }structures are described by the symmetry group
Fm$\overline{3}$m. When the A-cations and Z-waters occupy the interstitials in
an \textit{ordered} fashion the symmetry turns into F$\overline{4}$3m. The
F$\overline{4}$3m symmetry appears, for example, in CsFe[Cr(CN)6]$\cdot$%
H$_{2}$O \cite{Nomura}. It means in the ideal case that the system of
framework interstitials splits into two sublattices (1:1), as shown in Fig
1(a). In reality, each of these (1:1) sublattices is not represented by solely
A-ions and Z-waters, but occupied just unequally with these two species.

In the A-free structure, another sublattice may split into two subsystems that
is the anionic sublattice. When one proceeds from the disordered to ordered
placement of anionic vacancies the symmetry turns from Fm$\overline{3}$m to
Pm$\overline{3}$m giving birth to two dissimilar sublattices\cite{Buser}. In
the Pm$\overline{3}$m structure, two born sites are abundant with the ratio of
3:1. Four relevant lattice cells are shown in Fig.1 for the pairs of A-free
and A-filled structures.

Rietveld refinements were carried out using the above mentioned four
structural models and three symmetry groups known from the literature
\cite{Keggin,Ludi,Tokoro,Buser}. These treatments have shown that the most
symmetric Fm$\overline{3}$m model is sufficient to fit all the observed
reflections. When fitted with lower symmetries the XRD least-square analysis
has resulted in stagnant goodness of fit, unimproved, despite larger number of
fitting parameters. The intensity ratio $I_{200}/I_{220}$ changes dramatically
when K, Rb or Cs ions enter into the structure, as reported
previously\cite{RWZN}. The change becomes very large as the number of
electrons in an alcali ion increases, allowing for Cs a rough estimation of
its content even prior to Rietveld analysis.

The refined lattice parameters at ambient temperature are listed in Table I.
In all compounds, the lattice parameter is close to\ that of the low-spin
phase of CsFe[Cr(CN)$_{6}$]$\cdot$H$_{2}$O\cite{Nomura}, however, it is found
from magnetic measurements that the addressed in Table I compounds enclose the
Fe$^{2+}$ ions in the high-spin state that is unchanged down to 2 K.\ 

\subsection{Doublet and singlet orbital ground states of Fe$^{2+}$ in PB
analogues}

\subsubsection{Splitting the d-orbital energy level in anisotropic
environment}

In the fully A-filled PB analogues, the Fe$^{2+}$ ions are coordinated with
six nitrogens, however, at partial filling at least one C-water molecule
enters the first coordination sphere of Fe$^{2+}$. From the combined scheme of
compensating the charge misbalance between Fe$^{2+}$cations\ and Co(CN)$_{6}%
$]$^{3-}$ anions this is evidently the case of our samples (Table I).
Electronic configuration of a high-spin d$^{6}$ ion implies only one electron
extraneous with respect to isotropic half-closed d-shell composed of five
spins-up electrons. In an octahedral environment, the group of three t$_{2g}$
orbitals is available for this sixth electron. We show in Fig. 2 the
orientations of these orbitals relative to the water molecule.

In case of octahedral high-spin Fe$^{2+}$, the magnetic susceptibility of an
ion is contributed by the orbital motion of the t$_{2g}$ valence electron that
is added to the spin S=2 throgh the spin-orbit coupling. The valence
contribution of the electric field gradient (EFG) is predominant and exceeds
the lattice EFG term by an order of magnitude\cite{Nozik,Evans}. The t$_{2g}$
group is the source of valence term of EFG that is the main contribution of
the observed quadrupole splitting.
\begin{figure}
[ptb]
\begin{center}
\includegraphics[
height=2.9542in,
width=2.8029in
]%
{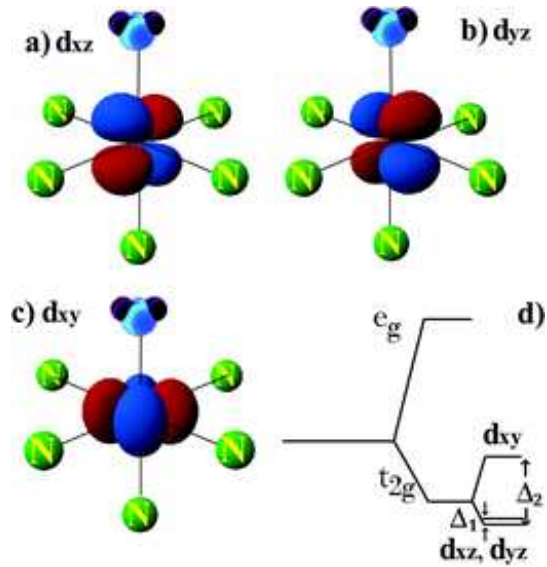}%
\caption{Formation of the orbital doublet ground state of Fe$^{2+}$ in\ the
"5+1" environment. In a Prussian Blue analogue, the valence spin-down electron
clouds are depicted for the Fe$^{2+}$ species coordinated by 5 nitrogens and
one water molecule. It is shown that two orbitals of the t$_{2g}$ group
(d$_{xz}$,d$_{yz}$)\ retain the equivalent energy (a,b), whereas the third one
(c) splits off upwards. A single spin-down electron occupies the
double-degenerate level forming the doublet orbital ground state.}%
\end{center}
\end{figure}

One of the orbitals of the \ t$_{2g}$ group in Fig. 2 (a,b,c) is perpendicular
to the direction pointing towards the water molecule. Corresponding energy
level would not be affected by the ligand replacement. The lobes of two other
orbitals (d$_{xz}$ and d$_{yz}$ in Fig.2) are at 45$^{\text{o}}$ with respect
to oxygen direction. Owing to reduced repulsion between oxygen and d$_{xz}$ or
d$_{yz}$ electrons corresponding levels would have lower energy. In the ground
state of Fe$^{2+}$ ion, these d$_{xz}$ and d$_{yz}$ orbitals share the sixth
d-electron (spin-down electron), while five spin-up d-electrons span
isotropically. The third (d$_{xy}$) orbital would split off its energy level
upwards from the ground state, formed by two degenerate d$_{xz}$ and d$_{yz}$
orbitals as shown in Fig. 2(d). The Fig. 2 (d) is relevant to zero spin-orbit
coupling. In case of combined effect of an axial field and of spin-orbit
interaction nine energy levels are produced, namely three singlets and six
doublets\cite{Konig1}.

\subsubsection{Magnetic properties of the high-spin Fe$^{2+}$ ions in the
distorted octahedral crystal field}

In the hexacyanocobaltates, where the high-spin Fe$^{2+}$ species are
surrounded by a low-spin complex anions [Co(III)(CN)$_{6}$]$^{3-}$, the
Fe$^{2+}$ site is a single source of paramagnetism. Orbital motion
contributing to the magnetic moment of Fe$^{2+}$ is unfrozen in the extent
determined by the degree of bonding covalence. Ligand replacement affects the
degree of valent electron delocalization. Variation of magnetic moment of
Fe$^{2+}$ ions over the sample is expected caused by the effect of a number of
C-waters entering the first coordination sphere of Fe$^{2+}$. There appears a
distribution of the Fe$^{2+}$ ions over the number of the C-water ligands. For
the magnetic measurements the samples 4-6 of the Table I were selected. \ The
compounds RbFe[Co(CN)$_{6}$]$\cdot$H$_{2}$O and CsFe[Co(CN)$_{6}$]$\cdot
$H$_{2}$O were selected because they are fully A-filled, enclosing only the
Z-water, and showing the single-site M\"{o}ssbauer spectra. As a counterpart
from the opposite family the compound Fe[Co(CN)$_{6}$]$_{0.67}\cdot5$H$_{2}$O
was selected (the sample No. 6). A convenient method to display the results of
magnetic measurements is to plot the experimental curves $\mu_{\text{eff}}(T)$
fitted with theoretical ones \cite{Konig1,Nature,Konig2,Figgis}. Accordingly,
in Fig. 3, the experimental data (1) and (2) for $\mu_{\text{eff}}(T)$ are
matched to a single set of fitting parameters ($\delta,\lambda,\kappa$). The
data (3) are approximated by a mean squared $\mu_{\text{eff}}=\sqrt{A_{1}%
\mu_{\text{eff}}^{2}(1)+A_{2}\mu_{\text{eff}}^{2}(2)}$ with two magnetic
moments $\mu_{\text{eff}}^{{}}(1),$ $\mu_{\text{eff}}^{{}}(2)$ and abundances
$A_{1}$ , $A_{2}$ adjusted to the parameters of subspectra observed in
corresponding M\"{o}ssbauer spectra.%

\begin{figure}
[ptb]
\begin{center}
\includegraphics[
height=2.6066in,
width=3.3728in
]%
{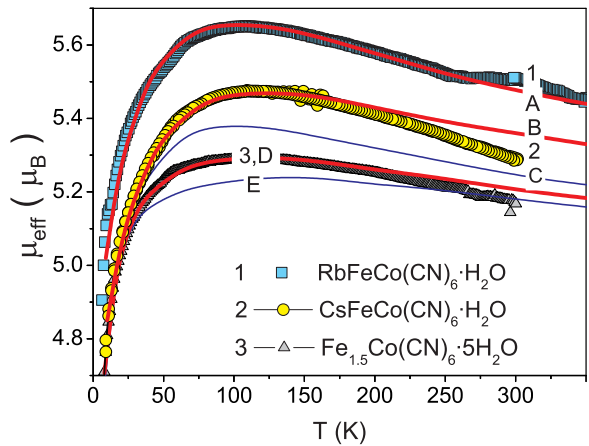}%
\caption{Experimental and theoretical curves of $\mu_{\text{eff}}$%
=2.83$\sqrt{\chi_{\text{mol}}T}$ vs. temperature for three compounds ( 4, 5
and 6 in Table I). The solid curves are from the theory\cite{Konig2} for the
parameters (A): $\Delta_{1}=190$ cm$^{-1}$, $\Delta_{2}=0$, $\lambda=63$
cm$^{-1},$ $\kappa=1$; (B) $\Delta_{1}=210$ cm$^{-1}$, $\Delta_{2}=0$,
$\lambda=70$ cm$^{-1},$ $\kappa=0.8$; (C) $\Delta_{1}=170$ cm$^{-1}$ ,
$\Delta_{2}=0$, $\lambda=55$ cm$^{-1},$ $\kappa=0.7$; (D) D=$\sqrt{0.4\cdot
C^{2}+0.6\cdot E^{2}}$; (E) $\Delta_{1}=\Delta_{1}=350$ cm$^{-1}$ ,
$\lambda=70$ cm$^{-1},$ $\kappa=0.6.$}%
\end{center}
\end{figure}

From the theory of magnetic properties of transition metal ions in
low-symmetry ligand field\cite{Konig1,Nature,Konig2,Figgis} the $\mu
_{\text{eff}}(T)$ is predicted to culminate around 100 K exceeding the
spin-only value of 4.9$\mu_{\text{B}}$ and depending on the crystal-field
splitting energy, the energy of spin-orbit coupling $\lambda$, and the orbital
reduction factor $\kappa$. The results tabulated in the reports of two
groups\cite{Konig2,Figgis} coincide\cite{Note}.

The curves $\mu_{\text{eff}}(T)$ can be described in terms of low-temperature
dip (below maximum), the negative slope (above maximum), and the overall
level. The width of the low-$T$ dip in the $\mu_{\text{eff}}(T)$ curves scales
with $\lambda$, therefore, the value of $\lambda$ can be most accurately
fitted, using the accurate claim that we can exclude mixing antiferromagnetism
into such a low-$T$ dip. On the other side, the negative slope of the
$\mu_{\text{eff}}$ vs. $T$ curve is steeper when the orbital doublet is lowest
($\Delta_{2}=0$). In this case (Rb and Cs-samples), large slope absolute value
evidences the orbital doublet ground state. The slope is almost unvaried with
$\Delta_{1}$, while the level of $\mu_{\text{eff}}$ depends strongly on
$\Delta_{1}$. However, the level of $\mu_{\text{eff}}$ depends also strongly
on $\kappa$. The correlation between these two dependencies reduces much the
accuracy of determination of $\Delta_{1}.$ Nevertheless, the choice between
the cases of $\Delta_{2}=0$ and $\Delta_{1}=\Delta_{2}$ is unambiguous.

With increasing the degree of t$_{2g}$ electron delocalization both values of
$\lambda$ and $\kappa$ are reduced from their ideal free-ion values of
$\lambda_{0}=-100$ cm$^{-1}$ and $\kappa_{0}=1$, respectively\cite{Figgis}. A
concommitent reduction of $\lambda/\lambda_{0}$ and $\kappa/\kappa_{0}$ in a
series of coordination compounds was evidenced by Figgis et al\cite{Figgis}.
Our results in Fig. 3 confirm such a correlated reduction, except in Rb-based
compound. However, if one assume that the molar weight of our sample is
slightly reduced by Rb-deficiency, the value of $\kappa$ would be reduced as
well. The deficiency of 0.06 (three times smaller than in our K$_{0.8}%
$Fe$_{1.1}$[Co(CN)$_{6}$]$\cdot$H$_{2}$O sample) would correspond to the
reduction of $\kappa$ from 1 to 0.8. In the samples 5 and 6, the observed
$\mu_{\text{eff}}(T)$ decreases with increasing temperature even more steeply
than does the maximally steep theoretical curve (Fig. 3). This result can be
explained\cite{Konig1} by the change of the covalence degree with temperature
as the overlap between the iron(II) 3d- and ligand charge clouds changes owing
to thermal expansion\cite{Konig1}.

\subsubsection{Temperature dependence of quadrupole splitting in A-filled
cobalticyanides (A=Na,K):}

We observe two doublets, both of them having the temperature-dependent
quadrupole splitting. In addition, the ratio of doublet areas is also strongly
temperature-dependent. The values of chemical shifts for both doublets in Fig.
4 are in the range 1.1 mm/s to 1.15 mm/s allowing us to identify both doublets
with the high-spin Fe$^{2+}$. A tiny spectral asymmetry indicates a trace of
oxidized (Fe$^{3+}$) component.

The external doublet is dominating at low temperatures. As the temperature is
raised an inverted doublet area ratio is observed. M\"{o}ssbauer spectra
suggest that the triple degeneracy of t$_{2g}$ is lifted over either in favor
of the orbital singlet or in favor of orbital doublet ground states. The
charge distribution is less anisotropic when the spin-down electron is shared
between two orbitals as shown in Fig. 2. We ascribe such a charge state to the
internal quadrupole doublet of Fig.4.

Larger charge anisotropy originates from localizing the sixth electron of
Fe$^{2+}$ on a single $d$-orbital (external QS doublet). From the QS ratio the
valence EFG term is expected to be twice larger for the orbital singlet than
for the orbital doublet. The literature overview confirms indeed that in
Fe$^{2+}$ the orbital singlet produces twice larger quadrupole splitting than
the orbital doublet ground
state\cite{Dezsi,Coey,Reiff1973,Latorre,Latorre3,Latorre2,Volland2,Volland,Brunot,Isotope,Hoy,Evans,Ganiel,Ingalls}%
.%

\begin{figure}
[ptb]
\begin{center}
\includegraphics[
height=4.4469in,
width=3.5241in
]%
{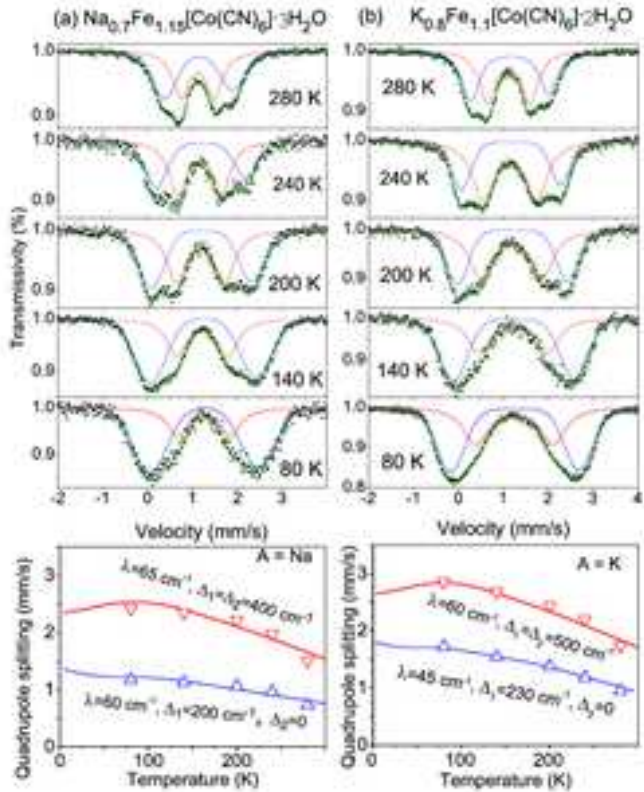}%
\caption{M\"{o}ssbauer spectra of the A-filled ferrous hexacyanocobaltates for
A=Na and K in temperature range between 80 K and 280 K. In the bottom panels,
the experimental temperature dependences of the quadrupole splitting
determined from least-squares fitting the M\"{o}ssbauer spectra \ are shown by
triangle symbols. The solid lines represents $\Delta E_{\text{Q}}(T)$ fitted
using the reduction factors $F^{\prime}(T)$ from Ref. \cite{Ingalls} with the
parameters values of $\lambda$, $\Delta_{1}$ and $\Delta_{2}$ indicated. For
the sake of clarity the minor site (cf. 2 to 5\% in Tables II and III) is not
shown in the spectra partitioning. Plausible origin of this site occupied by
iron in oxidized form (Fe$^{3+}$) is discussed in Sec.III D.}%
\end{center}
\end{figure}

In the bottom of Fig. 4, we fitted the curves $\Delta E_{\text{Q}}(T)$ using
the approach developed by Ingalls\cite{Ingalls}. The crystal-field splitting,
the spin-orbital interactions, the lattice term and covalency effects are
taken into account in a step-by-step procedure\cite{Merrithew} allowing to
estimate the triplet t$_{2g}$ splitting energies $\Delta_{1}$, $\Delta_{2}$,
the spin-orbit coupling constant $\lambda$ scaled-down by covalency effects
from the free-ion value of 100 cm$^{-1}$, and the lattice term $\Delta
$E$_{\text{latt}}$. In the first step, the so-called preliminary
results\cite{Merrithew} were derived from fitting the curves $\Delta
E_{\text{Q}}(T)$ with the theoretical expression\cite{Ingalls, Merrithew}:%

\begin{align*}
&  \Delta E_{\text{Q}}(T)=\hspace{2.4in}\\[0.31in]
&  \Delta E_{\text{Q}}^{0}\frac{(1+e^{\frac{-2\Delta_{1}}{k_{\text{B}}T}%
}+e^{\frac{-2\Delta_{2}}{k_{\text{B}}T}}-e^{\frac{-\Delta_{1}}{k_{\text{B}}T}%
}-e^{\frac{-\Delta_{2}}{k_{\text{B}}T}}-e^{\frac{-(\Delta_{1}+\Delta_{2}%
)}{k_{\text{B}}T}})^{%
\frac12
}}{1+e^{\frac{-\Delta_{1}}{k_{\text{B}}T}}+e^{\frac{-\Delta_{2}}{k_{\text{B}%
}T}}}%
\end{align*}

\begin{equation}
\end{equation}

Here $\Delta E_{\text{Q}}^{0}=(4/7)e^{2}Q\langle r^{-3}\rangle$ is the
quadrupole splitting at zero temperature, which is for Fe$^{2+}$ as high as 4
mm/s \cite{Merrithew} or even 4.5 mm/s\cite{Ingalls}. The Eq.(1) shows that
the value of $\Delta E_{\text{Q}}^{0}$ is reached at $T=0$ if $\Delta_{1}%
$=$\Delta_{2}$ (orbital singlet) but only the value of $\Delta E_{\text{Q}%
}^{0}/2$ is reached at $T=0$ if $\Delta_{2}=0$ (orbital doublet). The
preliminary fitting results imply the orbital doublet state for the inner
M\"{o}ssbauer doublet and the orbital singlet state for the outer
M\"{o}ssbauer doublet. Next step of the fitting procedure consist of taking
into account the spin-orbit coupling $\lambda$. Covalence effects reduce the
value of $\lambda$ compared to the free-ion value of $\lambda_{0}\approx$ -100
cm$^{-1}$. In this approximation, the Boltzmann distribution of the valence
electron [Eq.(1)] over three levels must be replaced with that over nine
levels\cite{Konig1}. We used in\ Fig.4 the theoretical results reported in
Figs. 3 and 5 of the Ingall's work \cite{Ingalls}, assuming an axially
symmetric local distortions of the FeN$_{6}$ octahedra within the entirely
cubic structure.

Finally, it is tempting to take into account the temperature-independent
lattice term. Most frequently the lattice term is opposite in sign to the
valence term and smaller by an order of magnitude\cite{Nozik}. If the lattice
term is subtracted from valence term, the overal $\Delta E_{\text{Q}}(T)$
slope would decrease, thus producing the underestimate of the crystal-field
splitting. The contribution $\Delta E_{\text{Q}}^{\text{latt}}$ of -10\% of
$\Delta E_{\text{Q}}^{0}$ leads to the underestimate of $\Delta_{1}$ of the
order of $\lambda$. Lacking the calculation of lattice sums we assumed $\Delta
E_{\text{Q}}^{\text{latt}}=0$ in Fig.4, however, if the ratio $\Delta
E_{\text{Q}}^{\text{latt}}/\Delta E_{\text{Q}}^{0}$ is indeed $\approx-0.1$,
then the expermental $\Delta E_{\text{Q}}(T)$ curves may be steeper than the
theoretical ones. The divergence of this type is observed in all $\Delta
E_{\text{Q}}(T)$ curves of Fig. 4. At approaching 300 K the experimental
$\Delta E_{\text{Q}}(T)$ curves become very steep. There could exist, however,
other source of such behavior. Yet another possibility is the steep change
$\Delta E_{\text{Q}}(T)$ in a proximity of a phase
transition\cite{Brunot,Isotope}, to be sought in these compounds presumably
above 300 K. Also, one may attribute the narrow doublet to a large fraction of
the supercooled high-temperature phase supposing the high- and low-temperature
phases isostructural. Indeed, from XRD no evidence of biphasic character of
samples were seen. However, if the transition is glasslike in its nature, a
single phase with disordered cubic structure may be hosting the residual
spatial inhomogeneity that implies the mixed orbital ground states coexisting
in a way by which different local structures coexist in a glass.

Thermally-induced change of areas of two M\"{o}ssbauer doublets shows that the
population of orbital doublet states grows with increasing temperature. It
appears that different local distortions substitute for one another with
preserving the global cubic symmetry. This gives us a reason to believe that
the coexisting orbital states originate from the \textit{local} distortions of
the FeN$_{6}$ octahedra that may have either static or dynamic character. In
the low-temperature region, the orbital singlet is lowest because the specific
distortion of the FeN$_{6}$ octahedra might be dynamically more stable. With
increasing temperature these polyhedra repose more time in a state of converse
distortion. The latter hosts the more isotropic distribution of the Fe$^{2+}$
valence electrons. Dynamic equilibrium shifts towards the reduced electronic
anisotropy with increasing temperature. Therefore, the orbital doublet states
appear to be partly thermally-induced.

{\small Table II. Parameters of M\"{o}ssbauer spectra (Fig. 4) in Na}$_{0.7}%
${\small Fe}$_{1.15}${\small [Co(CN)}$_{6}${\small ]}$\cdot3${\small H}$_{2}%
${\small O:\ isomer shift (IS) }$\delta${\small , quadrupole splitting (QS)
}$\Delta E_{\text{Q}}${\small , linewidths }$\Gamma${\small (mm/s), and
spectral abundance (\%).\allowbreak}

\qquad%
\begin{tabular}
[c]{|c|c|c|c|c|}\hline
{\small T (K)} & $\delta${\small (mm/s)} & ${\small \Delta E}_{\text{Q}}$ &
$\Gamma${\small (mm/s)} & {\small \%}\\
&  & {\small (mm/s)} &  & \\\hline
& {\small 1.142(2)} & {\small 1.56(1)} & {\small 0.48(1)} & {\small 44}\\
$280${\small K} & {\small 1.110(2)} & {\small 0.82(1)} & {\small 0.46(1)} &
{\small 51}\\
& {\small 0.46(1)} & {\small 0.3(2)} & {\small 0.27(2)} & {\small 5}\\
&  &  &  & \\
& {\small 1.161(8)} & {\small 2.06(2)} & {\small 0.50(2)} & {\small 40}\\
{\small 240 K} & {\small 1.168(2)} & {\small 0.99(3)} & {\small 0.51(2)} &
{\small 54}\\
& {\small 0.35(3)} & {\small 0.5(2)} & {\small 0.4(1)} & {\small 6}\\
&  &  &  & \\
& {\small 1.209(3)} & {\small 2.25(1)} & {\small 0.54(2)} & {\small 50}\\
{\small 200 K} & {\small 1.169(4)} & {\small 1.19(1)} & {\small 0.56(2)} &
{\small 46}\\
& {\small 0.4(1)} & {\small 0.5(2)} & {\small 0.4(4)} & {\small 3}\\
&  &  &  & \\
& {\small 1.218(2)} & {\small 2.467(9)} & {\small 0.67(1)} & {\small 65}\\
{\small 140 K} & {\small 1.240(3)} & {\small 1.287(1)} & {\small 0.51(1)} &
{\small 31}\\
& {\small 0.5(1)} & {\small 0.5(2)} & {\small 0.5(3)} & {\small 4}\\
&  &  &  & \\
& {\small 1.236(6)} & {\small 2.63(3)} & {\small 0.69(2)} & {\small 62}\\
{\small 80 K} & {\small 1.283(8)} & {\small 1.47(3)} & {\small 0.58(3)} &
{\small 35}\\
& {\small 0.5(1)} & {\small 0.5(2)} & {\small 0.4(2)} & {\small 3}\\\hline
\end{tabular}

Orbital-ground-state reversal in high-spin Fe$^{2+}$ compounds has been a
well-known phenomenon since the pioneering work of D\'{e}zsi and
Keszthelyi\cite{Dezsi}, in which a pair of doublets was observed in
Fe(ClO$_{4}$)$_{2}\cdot$6H$_{2}$O within a rather narrow temperature interval.
It was argued in this case, that two different orbital ground states coexists
within the transition range owing to the first-order phase
transition\cite{Coey}. In this system, the transition was observed by
calorimetry, neutron scattering and magnetic susceptibility
measurements\cite{Mikuli,neutron1,sus}. Order-disorder transformation in the
remote perchlorate ion bonding system built up of hydrogen bonds was
speculated to underlie the crystallographically undetected second-order
transition switching the FeO$_{6}$ octahedra from compression to elongation
along the trigonal axis\cite{Reiff1973}. In the isostructural fluoroborate
hydrate, Brunot\cite{Brunot} observed not so precipitous change in hydrogen
bond dynamics and deduced an inference of a different transition mechanism.
The temperature- and pressure-driven orbital ground state reversal phenomena
were proven to occur at the first-order phase transitions in iron(II)
fluorosilicate hydrate\cite{Volland2,Volland}. In thiourea high-spin Fe(II)
compound, no direct evidence of phase transition was found, while singlet and
doublet orbital ground states were found to coexist in a wider temperature
range \cite{Latorre,Latorre3,Latorre2}. The authors ruled out the
interpretation based on classical phase transition after they have conducted
the kinetic study\cite{Latorre3,Latorre2} that allowed them to specify the
transforming system as an intermediate between fully homogeneous and biphasic heterogeneous.

\bigskip

{\small Table III. Parameters of M\"{o}ssbauer spectra (Fig. 4) in K}$_{0.8}%
${\small Fe}$_{1.1}${\small [Co(CN)}$_{6}${\small ]}$\cdot2${\small H}$_{2}%
${\small O:\ isomer shift (IS) }$\delta${\small , quadrupole splitting (QS)
}$\Delta E_{\text{Q}}${\small , linewidths }$\Gamma${\small (mm/s), and
spectral abundance (\%).\allowbreak}

\qquad%
\begin{tabular}
[c]{|c|c|c|c|c|}\hline
{\small T (K)} & $\delta${\small (mm/s)} & ${\small \Delta E}_{\text{Q}}$ &
$\Gamma${\small (mm/s)} & {\small \%}\\
&  & {\small (mm/s)} &  & \\\hline
& {\small 1.135(2)} & {\small 1.76(1)} & {\small 0.44(1)} & {\small 44}\\
{\small 280K} & {\small 1.098(2)} & {\small 1.007(7)} & {\small 0.468(8)} &
{\small 54}\\
& {\small 0.4(1)} & {\small 0.3(2)} & {\small 0.4(2)} & {\small 2}\\
&  &  &  & \\
& {\small 1.173(2)} & {\small 2.207(8)} & {\small 0.38(2)} & {\small 36}\\
{\small 240 K} & {\small 1.152(2)} & {\small 1.185(9)} & {\small 0.59(1)} &
{\small 62}\\
& {\small 0.4(1)} & {\small 0.5(2)} & {\small 0.4(1)} & {\small 2}\\
&  &  &  & \\
& {\small 1.204(3)} & {\small 2.45(1)} & {\small 0.47(2)} & {\small 44}\\
{\small 200 K} & {\small 1.182(5)} & {\small 1.19(1)} & {\small 0.62(2)} &
{\small 54}\\
& {\small 0.4(2)} & {\small 0.5(2)} & {\small 0.6(4)} & {\small 2}\\
&  &  &  & \\
& {\small 1.234(3)} & {\small 2.68(2)} & {\small 0.63(2)} & {\small 62}\\
{\small 140 K} & {\small 1.229(7)} & {\small 1.54(3)} & {\small 0.60(3)} &
{\small 37}\\
& {\small 0.5(2)} & {\small 0.5(2)} & {\small 0.7(3)} & {\small 1}\\
&  &  &  & \\
& {\small 1.260(6)} & {\small 2.86(2)} & {\small 0.65(2)} & {\small 60}\\
{\small 80 K} & {\small 1.253(1)} & {\small 1.72(2)} & {\small 0.66(3)} &
{\small 39}\\
& {\small 0.5(2)} & {\small 0.5(2)} & {\small 0.7(3)} & {\small 1}\\\hline
\end{tabular}

\bigskip

In the same way, we avoid interpreting the data on Fig. 4 as a classical
thermodynamic first-order or second-order phase transition such as phonon-mode
condensation. In the compounds of the PB family, the ubiquitous cubic porous
framework allows for variable site and cell symmetries to be accommodated (cf.
structures in Fig.1). As the steric hindrances released in low-density
frameworks the coherence in atomic displacements ceases to exist. Additional
factor preventing the cooperativity of the local distortions is the disorder
related to incomplete A-filling.

As temperature decreases we observe in Tables II and III that the linewidths
of both major doublets increases. This is because each of the M\"{o}ssbauer
doublets is contributed by several Fe$^{2+}$ sites with different composition
of the 1$^{st}$ coordination sphere and with difference of positioning
relative to each other of the N and H$_{2}$O ligands. Depending on the number
and isomeric configuration (trans, cis) of the H$_{2}$O ligands around
Fe$^{2+}$ site the site contributes either to orbital doublet or to orbitals
singlet ground states. Two M\"{o}ssbauer doublets correspond to two types of
the ground state. Therefore, the values of parameters displayed on Fig. 4
($\Delta_{1},\Delta_{2},\lambda$) have the meaning of averages over the
orbital singlet and orbital doublet ground states. Subtle differences between
the crystal-field and spin-orbit parameters of the particlar environments give
rise to the low-temperature linewidth broadening of both doublets. Next, we
classify all the possible ligand configurations by the manner in which the
d-orbitals are filled depending on the symmetry of environment.

\subsubsection{Classification scheme of the Fe$^{2+}$ sites by the symmetry of
environment}

Our Rietveld refinements in both A-filled and A-free series of cobalticyanides
have revealed that their structures are best described by the symmetry
Fm$\overline{3}$m, allowing for the Fe$^{2+}$ ion to have a number of
different coordination environments. The Fe$^{2+}$ sites can be classified by
the number and configuration of N-ligands replaced with H$_{2}$O. In these
Fe$^{2+}$ sites, the lowest in energy state depends on the H$_{2}$O inclusion
in first coordination sphere, however, when H$_{2}$O is not included, the
ground state is determined by the local distortion. Owing to so-called
distortion isomerism\cite{Latorre} the ground state becomes not uniquely
defined. Shown in Fig.5 are the cases of distortion isomerism (a), orbital
singlet (b), and orbital doublet (c).%

\begin{figure}
[ptb]
\begin{center}
\includegraphics[
height=3.8614in,
width=3.2889in
]%
{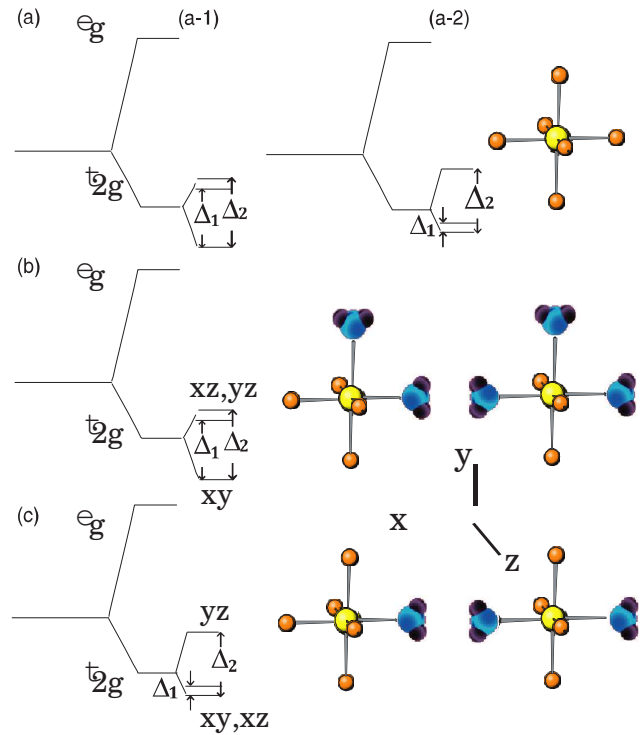}%
\caption{Energy level diagrams and corresponding coordinations for the orbital
singlet (a-1) and (b) and orbital doublet (a-2) and (c) ground states of the
Fe$^{2+}$ ion in ferrous hexacyanocobaltate. In a-1 (a-2) the diagram of level
splitting for the distorted FeN$_{6}$ ocahedron is shown for the elongation
(compression) along tetragonal (trigonal) axis.}%
\end{center}
\end{figure}

The distortion isomerism is the phenomenon closely related to the inability
for the sixth (spin-down) electron to be distributed isotropically over the
triply degenerate t$_{2g}$ orbital. Such an overmuch triple degeneracy is
universally lifted over by a local distortion of one or another type. In a
FeN$_{6}$ octahedron, the compression along tetragonal axis would stabilize
the orbital singlet lowest (a-1), while elongation along tetragonal axis
stabilizes the orbital doublet as a ground state (a-2). Opposite effects are
expected for the compression and elongation along the trigonal axis of the
octahedron\cite{Latorre2}. Self-coupling between electron localization and
local distortion eliminates the case of zero EFG valence term that is never observed.

This physical picture is a key to understanding the temperature variation of
the abundance of two M\"{o}ssbauer doublets in Fig. 4. Assuming that the
6N-coordinated Fe$^{2+}$ sites are contributing to both doublets, we observe
that the relative abundance of the orbital doublet and orbital singlet states
is varying with temperature owing to varied lattice distortion. Matsuda et
al\cite{Matsuda} have reported that the PB lattice possess an intrinsic
instability with respect to rotations of the rigid units of the [Co(CN)$_{6}%
$]$^{3-}$-type. The rotational displacements generate spontaneously to
stabilize the excited charge-transferred states, which arises owing to strong
lattice-electron coupling. The same rotational displacements cause also the
negative thermal expansion\cite{Matsuda}. Recently, the thermal expansion in
Fe$_{3}$[Co(CN)$_{6}$]$_{2}\cdot n$H$_{2}$O was reported to be indeed
negative, at least above 200 K\cite{Adak}.

We observe thus that the orbital ground state reversal accompanies the
distortion isomerism. One or another of coexisting distortion isomers prevails
in a particular temperature range. Largest charge anisotropy occurs at lower
temperatures owing to localizing the sixth electron of Fe$^{2+}$ on a single
$d$-orbital. The charge distribution is less anisotropic at higher
temperatures when the spin-down electron is shared between two orbitals. The
temperature variation of the abundance of two M\"{o}ssbauer doublets is very
smooth as the distortions are local and broadly distributed.

With decreasing the content of A-ions the anionic vacancies (water-filled
pores) are generated for charge compensation. When the pore concentration is
small only populated are the Fe$^{2+}$ sites which contain one H$_{2}$O ligand
as shown in Figs. 2 and 5(c). Two pores in a vicinity of a Fe$^{2+}$ species
occur either in \textit{trans-} or in \textit{cis-} configurations. Consider
in Fig. 5(c), for example, the ground state of the \textit{trans-} Fe$^{2+}$.
If we denote by $x$ the direction from the central Fe$^{2+}$ ion towards two
\textit{trans-} H$_{2}$O ligands then the lobes of both $xy$ and $xz$ orbitals
would be pointing to directions having an angle of 45$^{o}$ with the $x$ axis.
The lobes of the third t$_{2g}$ orbital ($yz$) would be pointing to the
direction orthogonal to the $x$-axis. Therefore the ground state of the
Fe$^{2+}$ ion in such FeN$_{4}$(H$_{2}$O)$_{2}$ octahedron would be the
orbital doublet. This is the same ground state as shown in Fig. 2 for the
Fe$^{2+}$ ion in the FeN$_{5}$H$_{2}$O octahedron with single water in first
coordination sphere. Orbital singlet in Fig. 5(b) is the ground state
corresponding to the \textit{cis}-configuration of FeN$_{4}$(H$_{2}$O)$_{2}$
and to the octahedron with three in-plane waters.

The H$_{2}$O placement into the first coordination sphere of Fe$^{2+}$
suppresses the distortion isomerism and stabilizes a singular ground state of
Fe$^{2+}$. In our samples Na$_{0.7}$Fe$_{1.15}$[Co(CN)$_{6}$]$\cdot3$H$_{2}$O
and K$_{0.8}$Fe$_{1.1}$[Co(CN)$_{6}$]$\cdot2$H$_{2}$O the density of anionic
vacancies is rather small and the population of the Fe$^{2+}$ ions surrounded
by two vacancies is also small. In these samples the inner doublet comes from
the Fe$^{2+}$ orbital doublet ground state associated with the FeN$_{5}$%
H$_{2}$O octahedron and with one of distortion isomers of FeN$_{6}$. The outer
doublet comes mainly from another distortion isomer of FeN$_{6}$ having the
orbital singlet ground state.

\subsubsection{A-filled cobalticyanides (A=Rb,Cs)}

Another manifestation of distortion isomerism is the difference between the
M\"{o}ssbauer spectra of the A-filled cobalticyanides obtained\ for Rb and Cs
by two different procedures. Using the standard technique, the A-filled
cobalticyanides were obtained by precipitation method. These compounds have
demonstrated the single-component M\"{o}ssbauer spectra\ (a) and (b) in Fig.6,
identified with orbital doublet ground state. On the other hand, when the
method of ionic exchange was applied to replace K by Cs in KFe[Co(CN)$_{6}%
$]$\cdot$H$_{2}$O, the second QS subspectrum, identified with the orbital
singlet, did not disappear. On the contrary, we observed the strong
enhancement of the outer doublet with $\Delta E_{\text{Q}}$=1.8 mm/s, from
32\% to 64\%. In addition, there appears a component (13\%) with extremely
large $\Delta E_{\text{Q}}$ of 3 mm/s. It is assigned to another orbital
singlet state with larger energy level splitting parameters $\Delta_{1}$,
$\Delta_{2}$. The difference between two external doublets of the sample (d)
is a manifestation another type of distortion isomerism unrelated to switching
between different orbital ground states\cite{Reiff}.%

\begin{figure}
[ptb]
\begin{center}
\includegraphics[
height=4.312in,
width=3.429in
]%
{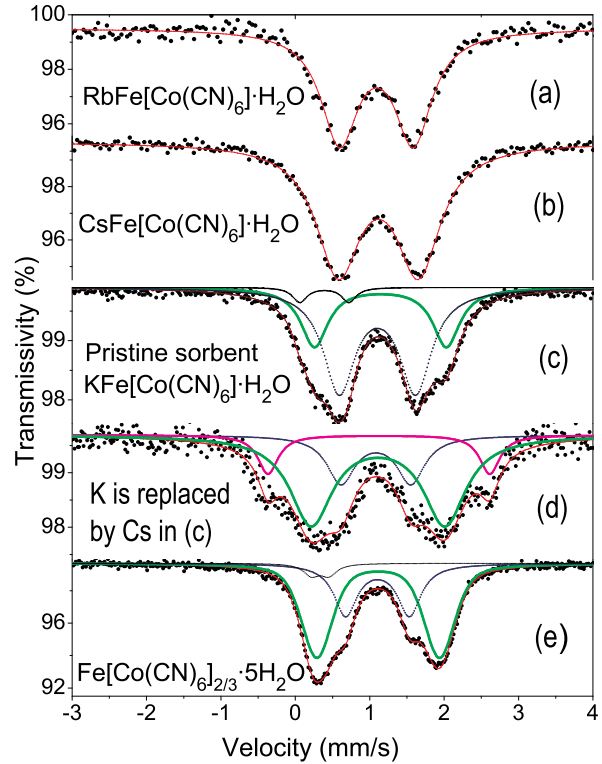}%
\caption{ M\"{o}ssbauer spectra of the A-filled ferrous hexacyanocobaltates
obtained in three different procedures: precipitation (a-c,e), and
ionic-exchange (d). The spectra fitting parameters are listed in Table II.}%
\end{center}
\end{figure}

Rietveld refinements resulted in similar compositions for two different
syntheses. The intensity of x-ray diffraction has very similar patterns,
except linewidth difference. Larger-size crystallites of CsFe[Co(CN)$_{6}%
$]$\cdot$H$_{2}$O are obtained via ionic exchange. In this method, the crystal
size of CsFe[Co(CN)$_{6}$]$\cdot$H$_{2}$O is determined by the size of the
KFe[Co(CN)$_{6}$]$\cdot$H$_{2}$O precursor crystals. The target compound
CsFe[Co(CN)$_{6}$]$\cdot$H$_{2}$O with such large crystals could not be
obtained by the standard method of precipitation from a solute. Despite of
similarity of the x-ray patterns, the QS of the ion-exchanged and precipitated
samples of CsFe[Co(CN)$_{6}$]$\cdot$H$_{2}$O indicates that the valent
spin-down electron has a very different distributions over the orbitals of
t$_{2g}$ group (Table IV).

\textbf{Table IV. }Parameters of M\"{o}ssbauer spectra (Fig. 6) in
as-synthesized (a-c) and (e) and Cs-sorbed(d) hexacyanocobaltates:\ isomer
shift (IS) $\delta$, quadrupole splitting (QS) $\Delta E_{\text{Q}}$,
linewidths $\Gamma$(mm/s), and spectral abundance (\%).\allowbreak%

\begin{tabular}
[c]{|c|c|c|c|c|}\hline
Compound & $\delta$(mm/s) & $\Delta E_{\text{Q}}$ & $\Gamma$(mm/s) & \%\\
&  & (mm/s) &  & \\\hline
(a):RbFe[Co(CN)$_{6}$] & 1.091(5) & 0.992(8) & 0.60(1) & 100\\
$\cdot$H$_{2}$O &  &  &  & \\
(b):CsFe[Co(CN)$_{6}$] & 1.116(3) & 1.077(4) & 0.71(1) & 100\\
$\cdot$H$_{2}$O &  &  &  & \\
(c):KFe[Co(CN)$_{6}$] & 1.144(3) & 1.77(1) & 0.44(2) & 32\\
$\cdot$H$_{2}$O & 1.104(2) & 1.03(1) & 0.52(2) & 64\\
& 0.39(1) & 0.66(2) & 0.23(3) & 5\\
&  &  &  & \\
(d): ion-exchanged & 1.111(5) & 1.80(3) & 0.69(4) & 64\\
K for Cs in (c) & 1.084(9) & 0.94(3) & 0.45(4) & 23\\
& 1.123(8) & 2.98(2) & 0.33(4) & 13\\
&  &  &  & \\
(e):Fe[Co(CN)$_{6}$]$_{2/3}$ & 1.114(2) & 1.65(1) & 0.45(1) & 64\\
$\cdot5$H$_{2}$O & 1.104(2) & 0.86(1) & 0.38(1) & 31\\
& 0.33(1) & 0.23(2) & 0.26(3) & 5\\\hline
\end{tabular}

We observe in the standard (not ion-exchanged) compounds that only one type of
the octahedron distortion is preferred when the large-size A-ions occupy the
interstitials, and this type of distortion stabilizes the orbital doublet
lowest. On the other hand, in the case of A=K or A=Na the appearance of the
outer doublet comes into play. In this series of PB analogues, several related
aspects of structure and properties turn out to be A-size-dependent. Whereas
in the case Cs(Rb)-containing PB analogue compounds the Cs(Rb) ions are
situated in the centers of the interstitial cubes, in the case of some
K(Na)-containing compounds the K(Na) ions are reported to be eccentrically
located \cite{Widman}. The mobility of alkali metal ions in the channels of
the hexacyanometalate structure decrease dramatically with increasing the
alcali ionic radius\cite{Widman}. The irreversible fixation of the less mobile
ions ensures the Cs-selectivity for the cobalticyanide sorbents, allowing the
ionic exchange to be used in practice for $^{137}$Cs removal from radioactive
wastes\cite{Ayrault,Loos}.

Although rather large linewidths were observed in the y=1 compounds (cf. in
Table IV $\Gamma=0.6$ for A=Rb and $\Gamma=0.7$ for A=Cs), there is no doubts
in attribution of the whole M\"{o}ssbauer spectra to the orbital doublet
ground states. In these cases, the XRD linewidths associated with the
smallness of crystal size\cite{RWZN} and the M\"{o}ssbauer linewidths
correlate well to each other. The distribution of EFG tensor component values
that causes the line broadening arises from the small size of crystals. These
compounds contains only the Z-water molecules, therefore, they are closer to
anhydrous salts rather than to crystal hydrates wherein the crystallization
water helps the precipitation of larger crystals. Similarly large
M\"{o}ssbauer linewidths were observed previously by Kosaka et al\cite{Nomura}
in the high-spin phase of the analogous chromate compound CsFe[Cr(CN)$_{6}%
$]$\cdot$H$_{2}$O devoid of crystallization water as well.

\subsubsection{Alcali-free ferrous cobalticyanide}

The isomer shifts in the M\"{o}ssbauer spectrum of the A-free Fe[Co(CN)$_{6}%
$]$_{2/3}$ (Fig.6e) indicates predominantly (95\% in best sample) the
high-spin Fe$^{2+}$. \ The ground states of the major part of the Fe$^{2+}$
species (64\%) are the orbital singlet states. Second place in population is
taken by the orbital doublet states. Third minor doublet is assigned to
Fe$^{3+}$ according to its isomer shift and magnetic measurements. Refinements
of the structure of Fe$_{1.5}$[Co(CN)$_{6}$]$\cdot$H$_{2}$O were useful to
confirm the assignments of three Fe$^{2+}$ M\"{o}ssbauer subspectra to the
defined structural sites.%

\begin{figure}
[ptb]
\begin{center}
\includegraphics[
height=4.7798in,
width=2.9992in
]%
{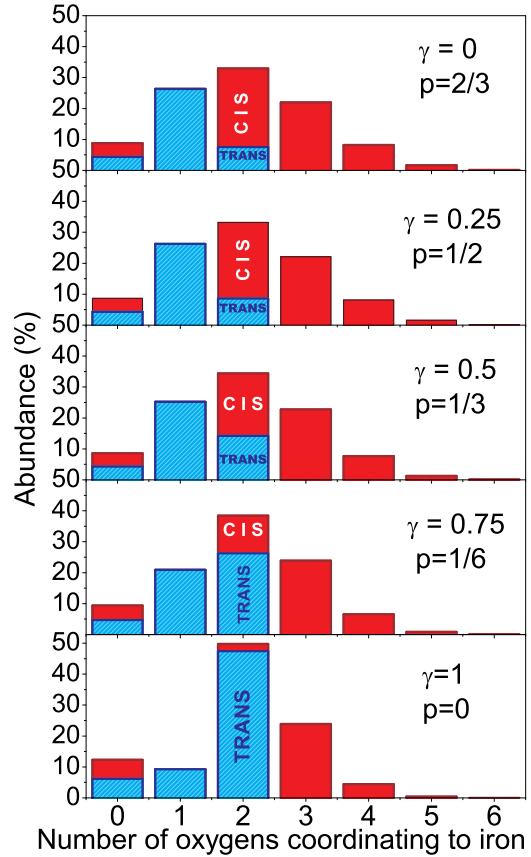}%
\caption{Evolution of the distribution of abundances of the coordination
polyhedra FeN$_{1-n}$O$_{n}$ ($n=0,1,2,3,4,5,6$) as the structure randomizes
starting from order parameter $\gamma=1$ (Pm$\overline{3}$m model, Fig1,b) to
$\gamma=0$ (Fm$\overline{3}$m model, Fig.1 d). Here $p=2(1-\gamma)/3$ is the
occupancy of the center of the cubic cell in Fig.1 (b). Total abundances of
orbital singlet and orbital doublet ground states are distinguished by color.}%
\end{center}
\end{figure}

Let us consider first the structure models of archetype Prussian Blue
investigated previously\cite{Buser,Herren}. The concentration of pores in the
A-free hexacyanocobaltate is very large, therefore, the pores cannot be
considered as isolated ones. All kinds of associations between the
water-filled pores underlie the resulting abundance of the Fe$^{2+}$ species
in various coordinations. Interactions between the pores at the stage of
synthesis may result in various distributions of the pores. For the Prussian
Blue Fe$_{4}$[Fe(CN)$_{6}$]$_{3}\cdot14$H$_{2}$O, depending on the growth
conditions, Buser et al\cite{Buser} could synthesize the single crystals with
varied degree of disorder. Structures of several crystals were refined using
the Fm$\overline{3}$m model, others were best refined using the Pm$\overline
{3}$m model. \textit{A priori}, fully randomized distribution of the pores is
consistent with the Fm$\overline{3}$m model (Fig. 1d). Low symmetry
(Pm$\overline{3}$m) becomes applicable when one finds unequal occupancies for
the cell body center and edge centers (Fig. 1b). These sites are abundant with
the ratio of 1:3. If $p$ is the occupancy of the body center, and $1-p/3$ is
the occupancy of the edge centers, the conditions $p=0$ and $p=3/4$ correspond
to fully ordered and random PB structures, respectively\cite{Buser}. In
ferrous hexacyanocobaltate, owing to different valence ratio the same
conditions correspond to $p=0$ and $p=2/3$. Intermediate partially-ordered
structures can be characterized by the order parameter $\gamma=1-3p/2$ .

As $\gamma$ decreases, the integral value of $A_{\text{D}}$ shows in Fig. 7
the percentages of 63\%, 52\%, 44\%, 39\% and 38\% for $\gamma=1,%
\frac34
,%
\frac12
,%
\frac14
,0$ , respectively. Main trend of reduction of $A_{\text{D}}$ is determined by
the reduction of the \textit{trans}-subpopulation of FeN$_{4}$(H$_{2}$O)$_{2}%
$. The orbital singlet ground state prevails in the disordered structure
described by the Fm$\overline{3}$m model. In this case ($\gamma=0$), the
expected area of the external doublet $A_{\text{S}}$\ reaches the value of
62\% similar to experimental result of 64\% in Table IV.

In Fig. 7, we have calculated the abundance of various coordinations of
Fe$^{2+}$ for several values of $\gamma$ between 0 and 1. The total abundance
of the Fe$^{2+}$species in the orbital doublet ground state $A_{\text{D}}$ is
made up mainly of the abundances of FeN$_{5}$H$_{2}$O and \textit{trans}%
-FeN$_{4}$(H$_{2}$O)$_{2}$. Relatively small number of water-free octahedra
FeN$_{6}$ is assumed to be halved between two distortion isomers.%

\begin{figure}
[ptb]
\begin{center}
\includegraphics[
height=3.5042in,
width=3.4281in
]%
{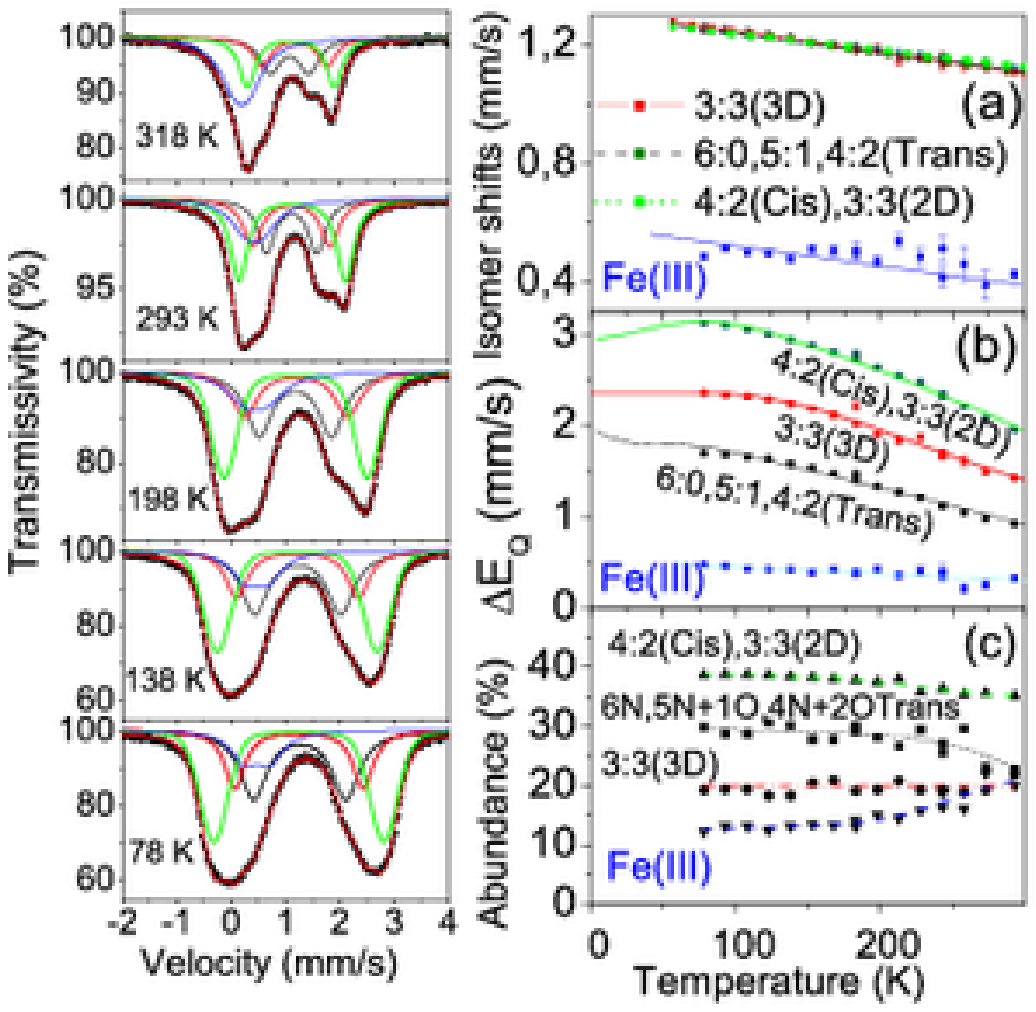}%
\caption{Left panel: M\"{o}ssbauer spectra of Fe[Co(CN)$_{6}$]$_{0.67}\cdot
$5H$_{2}$O at 78, 138,198, 293 and 318 K fitted with four quadrupole doublets.
The external (green color) doublet is assigned to Fe$^{2+}$ (orbital singlet)
inside the octahedra FeN$_{4}$(H$_{2}$O)$_{2}$ with two H$_{2}$O in
\textit{cis}-conformation and FeN$_{3}$(H$_{2}$O)$_{3}$ with three H$_{2}$O in
one plane (2D conformation). Middle doublet (red color) is assigned to the
orbital singlet ground state inside the FeN$_{3}$(H$_{2}$O)$_{3}$ polyhedra
with waters in 3D conformation. Inner Fe$^{2+}$ doublet is associated with the
FeN$_{6}$, FeN$_{5}$H$_{2}$O, and \textit{trans}-FeN$_{4}$(H$_{2}$O)$_{2}$
polyhedra all favoring the orbital doublet states. The spectra below 293 K are
reversible. Additional doublet (blue) has the isomer shift of Fe$^{3+}$. The
spectra below 293 K are reversible. The sample warmed up to 318 K (uppermost
spectrum) is changed irreversibly ( for further details see text in Sec. III
D). \ Right panel : temperature dependencies of the fit parameters of four
quadrupole doublets: isomer shifts (a), quadrupole spittings (b) and
abundances. Quadrupole splittings in (b) are fitted with the reduction factors
$F^{\prime}(T)$ from Ref. \cite{Ingalls}. The solid lines represents the
theoretical $\Delta E_{\text{Q}}(T)$\ for $\lambda=58$ cm$^{-1}$, $\Delta_{1}$
=$\Delta_{2}$ =320 cm$^{-1}$ for the external doublet attributed to the 4:2
(Cis) and 3:3 (2D) sites; $\lambda=80$ cm$^{-1}$, $\Delta_{1}$ =$\Delta_{2}$
=300 cm$^{-1}$ for the intermediate doublet attributed to the 3:3 (3D) site$;$
$\lambda=45$ cm$^{-1}$, $\Delta_{1}$ =150 cm$^{-1}$ $\Delta_{2}$ =0 for the
internal doublet attributed to the 6:0, 5:1 and 4:2(Trans) sites.}%
\end{center}
\end{figure}

Abundance of the orbital singlet $A_{\text{S}}$ is composed \textit{cis}%
-FeN$_{4}$(H$_{2}$O)$_{2}$, FeN$_{3}$(H$_{2}$O)$_{3}$ and remaining FeN$_{6}%
$(distortion isomer). Both in-plane and 3D ligand configurations in the 3:3
octahedron FeN$_{3}$(H$_{2}$O)$_{3}$ contribute to $A_{\text{S}}$. First, when
3 oxygens coordinating to iron are in one plane the $x$, $y$ and $z$ axes
become all nonequivalent. In this case, the spin-down electron cloud loses the
axial symmetry, and the t$_{2g}$ group splits into three singlets. Second,
when three oxygens are in 3D configurations, all three axes $x$, $y$ and $z$
are equivalent; nevertheless, the cubic symmetry is lost. Since H$_{2}$O is a
weaker ligand than N this geometry can be viewed as an octahedron with three
ligands expelled to longer distance than other three. Equivalent spatial
distortion is the elongation along the trigonal axis. Such a distortion was
shown to stabilize the orbital singlet lowest\cite{Latorre}.

While there are just two sorts of the ground states, the inspection of Fig.7
for p=2/3 shows that there are eight sites of Fe$^{2+}$ having different
ligand configurations with the abundance of 5\% or more. Among the octahedra
with n=0,1,2,3 and 4 those with n=0, 2 and 3 are found in two isomeric
configurations: compression-elongation (n=0), cis-trans (n=2) and 2D-3D (n=3).
These eight sites are to be completed with ninth site if the oxidized
component (Fe$^{3+}$) is present in the sample. It appears that the most
sophisticated way of spectra evaluation consist of partitioning the spectral
area into nine doublets. Even if the method would reproducibly fit the
multiple spectra, it would be uneasy task to assign the resulting subspectra
to a particular isomer. Highly valuable remains the models in which less than
nine subspectra are assumed. In the A-free cobalticyanide (Fig.8), a plausible
model of 4 subspectra is constructed through breaking up the external doublet
with the excessive linewidth into two doublets. Resulting intermediate
quadrupole doublet is assigned to the 3-dimensional 3:3 configuration of the
ligands (3N+3H$_{2}$O). Corresponding crystal fields have a trigonal symmetry
axis. We estimate this symmetry to reduce slightly the ensuing quadrupole
splitting. Thus, in this model, the orbital singlet states stabilized in the
crystal fields of trigonal symmetry are distinguished from the orbital singlet
settled in the 4:2 (cis) and 3:3 (2D) environments.

In the experiment with enriched sample of $^{57}$Fe[Co(CN)$_{6}$]$_{0.67}%
\cdot5$H$_{2}$O the spectra were collected with the step $\Delta T=15$ K
(Fig.8) The evaluation of these spectra with five, and six doublets were also
attempted. These preliminary analyses showed that the model of four doublets,
while providing already strong improvement of $\chi^{2}$, ensures also the
most reproducible and stable solution. The model is changing its parameters
smoothly as temperature changes.

Orbital ground state reversal observed above in the A-filled compounds (Fig.5)
is not seen in the A-free ones (Fig.8). This is naturally explained by the
small contribution of the n=0 sites susceptible to the distortion isomerism.
It is shown in Fig.7 that the number of such sites does not exceed 8\%. Other
sites strongly perturbed by the CN ligand replacement exhibits persistent
ground states independent of temperature.

Parameters of crystal-field splitting and spin-orbit coupling derived from
fitting the temperature dependencies of $\Delta E_{\text{Q}}$ are in agreement
with the averaged curve of $\mu_{\text{eff}}(T)$ on Fig. 3. The splitting of
340 cm$^{-1}$ estimated previously in Fe$_{3}$[Co(CN)$_{6}$]$_{2}\cdot5$%
H$_{2}$O by Rasmussen and Meyers\cite{Ras} corresponds closely to one of our
results, namely, to the crystal fields in the Fe$^{2+}$ sites contributing to
the external doublet.

\subsection{Models of short-range order in PB analogues}

The concentration of pores in archetypal Prussian Blue, ferric ferrocyanide
Fe[Fe(CN)$_{6}$]$_{1-\delta}\cdot$3.5H$_{2}$O, is $\delta=1/4$, so that the
ratio $1-\delta:\delta$ \ is 3: 1. The ordered Pm$\overline{3}$m structure
proposed for Prussian Blue in Ref. \cite{Buser} contains the high-spin cations
only in coordinations FeN$_{6}$ (25\%) and trans-FeN$_{4}$(H$_{2}$O)$_{2}$
(75\%). In ferrous hexacyanocobaltate, the ratio $1-\delta:\delta$ \ is 2: 1.
Therefore, for a vacancy-ordered phase of such a stoichiometry one cannot
expect the same set of coordinations as for the Pm$\overline{3}$m structure of
the Ref. \cite{Buser}.

In order to construct an ordered phase with the ratio $1-\delta:\delta$ \ of
2: 1 we try to fill 3 sites with Fe, 2 sites with Co, 1 site with Co
vacancies, and 0 sites with Fe vacancies. We find that among cubic structures
of NaCl-type there is no superstructure with exact site ratio 3:2:1:0,
however, a unique superstructure with the ratio of 30:20:12:2 exists. Using
the body-centered Im$\overline{3}$m model we can develop the proper
arrangement of the vacancies with a variable order parameter. Table V shows
the details of the fully ordered ($\gamma=1$) Im$\overline{3}$m structure. The
order parameter $\gamma=1-8p/3$ is defined via the occupancy $p$ of the (12e)
position that is the core of complex anion [Co(CN)$_{6}$]$^{3-}$.

\textbf{Table V. }Wyckoff positions, atom type, relative coordinates and
occupancies of the Na and Cl nodes in the superstructure Im$\overline{3}$m
model on the basis NaCl-type framework.%

\begin{tabular}
[c]{|llllll|}\hline
Wyckoff & Atom & x & y & z & Occ.\\
2a & Fe & 0 & 0 & 0 & 0\\
6b & Fe \textit{Trans} & 0 & 1/2 & 1/2 & 1\\
8c & Co & 1/4 & 1/4 & 1/4 & 1\\
12d & Co & 1/4 & 0 & 1/2 & 1\\
12e & Co & 1/4 & 0 & 0 & 0\\
24h & Fe \textit{Cis} & 0 & 1/4 & 1/4 & 1\\\hline
\end{tabular}

Among total 14 vacancies per cubic cell of the Im$\overline{3}$m structure
(Table V) we place 12 vacancies of the complex anion Co(CN)$_{6}$ and 2
cationic (Fe$^{2+}$) vacancies. We obtained the ordered structure of
Fe[Co(CN)$_{6}$]$_{2/3}^{\text{\textbf{.}}}\cdot\frac{85}{16}$H$_{2}$O which
contains 2 very large cross-shaped pores (6${\LARGE +}$1-pores, or
${\LARGE +}$-pores). Each cationic vacancy is surrounded by 6 anionic
vacancies. The hydration degree $\frac{85}{16}$ is calculated assuming 14
water molecules per standard pore and an additional water molecule per
Fe$^{2+}$ vacancy. Thus, the 6${\LARGE +}$1-pore is composed of 6
standard$\ $pores aggregated into 3D crosslike body. The edge of the
Im$\overline{3}$m cell is twice longer than the edge of the Pm$\overline{3}$m
cell and the cell volume is 8-fold.

The body-centered Im$\overline{3}$m model exhibits only two kinds of Fe$^{2+}$
sites with the abundance ratio of 3:1, similar to that of the Pm$\overline{3}%
$m model.\ In both models, the major site (75\%) is coordinated with two
waters. However, in the Pm$\overline{3}$m model the major site is coordinated
by waters in \textit{trans}-conformation, whereas in the Im$\overline{3}$m
model the major site is coordinated by waters in \textit{cis}-conformation.
The minor site in Pm$\overline{3}$m model is coordinated by six nitrogens, and
in Im$\overline{3}$m model by four nitrogens and two waters in \textit{trans}-conformation.

It is important that we were able to construct the superstructure model, which
strongly favors the \textit{cis}-conformation in agreement with prevailing
external doublet in M\"{o}ssbauer spectra. Although the random Fm$\overline
{3}$m model agrees as well with the experiment, the lack of any short-range
order between pores appears to be a severe limitation of this model because
the vacancies bearing a charge interact with each other repulsively. We must
note also that the luck of any superstructure reflections in diffraction
patters may not be an evidence of a missing short-range order, but only the
evidence of lacking \textit{long-range} order. The suggested\ Im$\overline{3}%
$m structure is a plausible model of the short-range order in the
mixed-valence 2:3 PB analogues.

Although the superstructure reflections, predicted by either Pm$\overline{3}$m
or Im$\overline{3}$m models, were not observed by x-ray diffraction, using the
probabilities of local structure patterns observed by such a local technique
as M\"{o}ssbauer spectroscopy one can identify the proper model of short-range
order. The Im$\overline{3}$m model predicts the strong predominance (3:1) of
\textit{cis}- over \textit{trans}- population among the Fe$^{2+}$
environments. The model describes the situation when concentration of
elementary vacancies is so high that they must be in small clusters. In
contrast, the Pm$\overline{3}$m model describes the limit when all vacancies
are still isolated, however, no more vacancies can be added without
clustering. If they are actually added to proceed from $\delta=1/4$ to
$\delta=1/3$ (as in the bottom panel of Fig.7), there will result the
extra-holey Pm$\overline{3}$m model that fails to describe the M\"{o}ssbauer
spectral data. In our experiments, we observe the prevailing abundance of the
\textit{cis-}conformations related to the extended crosslike pores. The
aggregation of small pores into extended pores underlies the large (up to 750
m$^{2}$/g) specific surface and unique gas sorption properties of these
materials\cite{Kaye}.

\subsection{ Charge-transferred states in the PB analogues}

The charge-transferred states which were discussed above to induce the
distortion isomerism are seen in the third small doublet with IS of Fe$^{3+}$
in Figs. 6 (c,e) and 8. A slight asymmetry appears also in spectra of Fig.4,
although these spectra can be just a little less well fitted without
introducing explicitly the Fe$^{3+}$ component. The pairs of Fe$^{2+}%
$/Co$^{3+}$ and Fe$^{3+}$/Co$^{2+}$ can be considered as valence tautomers,
and we explain the asymmetry in these spectra by electron transfer from
Fe$^{2+}$ to Co$^{3+}$. The formation and rupture of the chemical bonds
implies no migration of atoms but only migration of electrons. The cobalt ions
remain in low-spin state as the sixth electron of Fe$^{2+}$ transfers from
t$_{2g}$ orbital of Fe$^{2+}$ to e$_{g}$ orbital of Co$^{2+}$. Other compound
susceptible to valence tautomerism were known previously\cite{Tokoro,Gutlich}.
For example, the Prussian Blue analogue RbMn[Fe(CN)$_{6}$]$\cdot$H$_{2}$O,
when synthesized in Rb excess that fills enough framework interstitials with
Rb, crystallizes in the structure shown in Fig. (1a). In this system, the
tautomers Fe$^{2+}$/Mn$^{3+}$and Fe$^{3+}$/Mn$^{2+}$ are able to form
different crystal phases which give way to each other as temperature changes.

In the ferrous cobalticyanides, no abrupt phase change was found to be caused
by the charge-transferred states. On the other hand, we observed that the
content of Fe$^{3+}$ can be changed depending on the excess of FeCl$_{2}$ and
dilution degree in the course of precipitation of Fe[Co(CN)$_{6}$]$_{2/3}%
\cdot5$H$_{2}$O. Rapidly synthesized isotope-enriched compound contained up to
15-20\% of Fe$^{3+}$, however, the content of Fe$^{3+}$ was reduced down to
just 5\% through reducing the reaction rate in dilute solutions. Most
probably, the lack or misplacement of water molecules is needed to stabilize
locally the charge-transferred states and make them observable in
M\"{o}ssbauer spectra. It appears in this case that the rapidly prepared
samples are not saturated in C-water. The sites of Fe$^{2+}$ nearest to the
nodes lacking the coordination water are subject of local tautomer
transformation. In a slowly synthesized Fe$_{1.5}$[Co(CN)$_{6}$]$\cdot5$%
H$_{2}$O such crystallization defects are reduced.

Measuring the M\"{o}ssbauer spectra above room temperature supports this
concept. With raising temperature the area of the Fe$^{3+}$ subspectrum starts
to increase irreversibly. Thermogravimetric analysis showed that the loss of
sample weight occurs at heating practically starting from room temperature.
The M\"{o}ssbauer spectrum measured \textit{in situ} in a furnace at\ 318 K
shows the area of the Fe$^{3+}$ doublet increased up to 37\%. The same sample
measured \textit{ex-situ} after prolonged (1 week) heat treatment at 60$^{o}$C
showed the area of the Fe$^{3+}$ increased up to 57\%. The equilibrium
hydration degree at 60$^{o}$C is related with nearly half iron species
transformed to Fe$^{3+}$. Longer exposure in air at at 60$^{o}$C induces no
further increase of the spectral asymmetry and the area of spectral component
with IS of 0.39 mm/s remains at the level of 60\%.

The isomer shift of 0.39 mm/s is attributable, in principle, either to
high-spin Fe$^{3+}$ or to low spin Fe$^{2+}$. The latter was observed below
the spin-state transition in CsFe[Cr(CN)$_{6}$] \cite{Nomura}. Theoretical
approaches to lattice-dynamic calculations\cite{Wojdel},\cite{Middlemiss} show
that the change in the electronic state is closely related to the structure
expansion as the lattice parameter grows between the low-spin and high-spin
forms from 1.03 nm to 1.07 nm. In Fe$_{3}$[Co(CN)$_{6}$]$_{2}\cdot n$H$_{2}$O,
the thermal lattice expansion is negative\cite{Adak}. To verify that we do not
deal with a \textit{reverse} spin transition \cite{Hayami,Manoharan}, we have
compared the magnetic susceptibilities of the samples having 5\% and 57 \% of
the subspectrum with IS of 0.39 mm/s . In the heat-treated sample, the
susceptibility increased too little compared to what is expected for the
increase of $\mu_{\text{eff}}$ upto 5.9 $\mu_{\text{B}}$. With the weight loss
taken into account this result would rather indicate the formation of an
intermediate-spin state of Fe$^{3+}$ which is known for strongly distorted
octahedral coordinations with two alternate ligands in \textit{cis}%
-configuration\cite{Petra}.

\section{Conclusion}

M\"{o}ssbauer spectroscopy was applied to study the mixed-valence (Fe$^{2+}%
$-Co$^{3+}$) PB analogues which displayed the mixed-orbital ground states of
Fe$^{2+}$ ions. The ground state mixing originates from both the distortion
isomerism and the varied composition of Fe$^{2+}$ nearest-neighbor
environment. In the A-free PB analogues, water molecules enter the first
coordination sphere of the Fe$^{2+}$ ions. There appears two kinds of
anisotropic ligand configurations which are the intrinsic generant of the
singlet and doublet orbital ground states. The Fe$^{2+}$ sites are thus
classified into two types. 

Distribution of the anionic vacancies (water-filled pores) over the framework
is characterized by the order parameter for two long-range order models
Pm$\overline{3}$m or Im$\overline{3}$m. These models are relevant to the PB
analogues of the 4:3 and 3:2 stoichiometries, respectively. Two water
molecules within the first coordination sphere of Fe$^{2+}$ is the most
populated configuration of the random model Fm$\overline{3}$m and ordered
models Pm$\overline{3}$m or Im$\overline{3}$m. The short-range order event
occurrence in PB analogues of the 3:2 and 4:3 stoichiometries favors the cis-
and trans- conformations, respectively. Orbital singlet is the lowest in
energy state for the cis-conformation. Accordingly, in the A-free ferrous
cobalticyanide PB analogue, the M\"{o}ssbauer spectra of Fe$^{2+}$ are
dominated by the larger quadrupole-splitting component.

In the A-filled PB analogues, where the the water ligands are withdrawn from
the Fe$^{2+}$ coordination sphere, the least anisotropic charge distribution
attainable for the cases of A=Rb and Cs is the orbital doublet ground state.
For A=Na and K, the thermally-induced interconversions between the doublet and
singlet ground states were observed attributed to distortion isomerism.
Charge-transfered states intrinsic in the PB analogues with negative thermal
expansion\cite{Matsuda} are suggested to be the source of the lattice
distortions associated with the distortion isomerism. Asymmetry of
M\"{o}ssbauer spectra is attributed to the admixture of a small content (at
the level of $\sim$5\%) of the Fe$^{3+}$ species associated with the
charge-transferred states.

Since the octahedral sites FeN$_{6}$ possess a higher symmetry than the sites
FeN$_{6-n}$(H$_{2}$O)$_{n}$ we observe upon the insertion the A-ions into A
free compounds that the value of M\"{o}ssbauer QS drops in ferrous
cobalticyanides similarly to the archetypal Prussian Blue Fe$_{4}$%
[Fe(CN)$_{6}$]$_{3}\cdot14$H$_{2}$O\cite{RWZN}. However, when the large-size
A-ions (Cs, Rb) replace the small-size ones (Na,K) through ionic exchange we
observe the larger QS associated with lower symmetry local distortion isomer;
the sixth d-electron of the Fe$^{2+}$ cation is self-localized on a single
sublevel (orbital singlet) of the t$_{2g}$ group owing to strong lattice
distortions around the implanting sites of the low-mobility ions.

\section{\bigskip Acknowledgement}

This work was partly supported by the Chinese Academy of Sciences Visiting
Professorships for Senior International Scientists. Grant No. 2011T1G15.
Financial support obtained from the Chinese Academy of Sciences for
\textquotedblleft100 Talents\textquotedblright\ Project, the National Natural
Science Foundation of China (No. 11079036) and the Natural Science Foundation
of Liaoning Province (No. 20092173) is also greatly acknowledged.


\begin{thebibliography}{99}                                                                                               %


\bibitem {Kato}K. Kato, Y. Moritomo, M. Takata, M. Sakata, M. Umekawa, N.
Hamada, S. Ohkoshi, H. Tokoro, and K. Hashimoto, Phys. Rev. Lett. \textbf{91,}
255502 (2003).

\bibitem {GutlichCSR}Ph. G\"{u}tlich, Y. Garcia, and H.A. Goodwin, Chem. Soc.
Rev. \textbf{29}, 419-427 (2000).

\bibitem {LI}E. Coronado, M.C. Gim\'{e}nez-L\'{o}pez, T. Korzeniak, G.
Levchenko, F.M. Romero, A. Segura, V. Garc\'{\i}a-Baonza, J.C. Cezar, F.M.F.
de Groot, A. Milner. and M. Paz-Pasternak, J. Am. Chem. Soc. \textbf{130,
}15519-15532\textbf{ }(2008).

\bibitem {Chong}C. Chong, M. Itoi, K. Boukheddaden, E. Codjovi, A. Rotaru, F.
Varret, F.A. Frye, D.R. Talham, I. Maurin, D. Chernyshov, and M. Castro, Phys.
Rev. \textbf{B} \textbf{84,} 144102 (2011).

\bibitem {Koudr}A. B. Koudriavtsev, W. Linert, J. Struct. Chem. Engl. Transl.
\textbf{51,} 335-365 (2010).

\bibitem {Moritomo}Y. Moritomo, F. Nakada, H. Kamioka, T. Hozumi, and S.
Ohkoshi, Phys. Rev. \textbf{B 75}, 214110 (2007).

\bibitem {Kabir}M. Kabir and K.J. VanVilet, Phys. Rev. \textbf{B 85}, 054431 (2012).

\bibitem {Rykov}A.I. Rykov, Y. Ueda, K. Nomura, and M. Seto, Phys. Rev.
\textbf{B 79} 224114 (2009)\textbf{.}

\bibitem {Dezsi}I. D\'{e}zsi and L. Keszthelyi, Solid St. Comm. \ \textbf{4},
511 (1966).

\bibitem {Coey}J. M. D. Coey, I. D\'{e}zsi, P. M. Thomas, and P.J. Ouseph,
Phys. Lett. \textbf{41A}, 125 (1972).

\bibitem {Reiff1973}W.M. Reiff, R.B. Frankel, C.R. Abeledo, Chem. Phys. Lett.
\textbf{22}, 124 (1973).

\bibitem {Latorre}R. Latorre, C.R. Abeledo, R.B. Frankel, J.A. Costamagna,
W.M. Reiff, E. Frank, J. Chem. Phys. \textbf{59,} 2580-2585(1973).

\bibitem {Latorre3}R. Latorre, J.A. Costamagna, E. Frank, C.R. Abeledo, and
R.B. Frankel, J. Physique (Paris) \textbf{C6}, 635 (1974).

\bibitem {Latorre2}R. Latorre, J.A. Costamagna, E. Frank, C.R. Abeledo, and
R.B. Frankel, J. Inorg. Nucl. Chem. \textbf{41}, 649-655 (1979).

\bibitem {Volland2}U. Volland, S. H\"{o}sl, H. Spiering, I. D\'{e}zsi, T.
Kem\'{e}ny, and D.L. Nagy, Solid St. Comm. \textbf{27},49 (1978).

\bibitem {Volland}U. Volland, J. Phys. Colloques, \textbf{41}, C1-309-C1310 (1980).

\bibitem {Brunot}B. Brunot, Chem. Phys. Lett. \textbf{29}, 371 (1974).

\bibitem {Isotope}B. Brunot, Chem. Phys. Lett. \textbf{29}, 368 (1974).

\bibitem {Hoy}G.R. Hoy, and F. de. S. Barros, Phys. Rev. \textbf{139},
A929-A934 (1965).

\bibitem {Ras}P.G. Rasmussen and E.A. Meyers, Polyhedron \textbf{3} 183-190 (1984).

\bibitem {Reg}E. Reguera, H. Yee-Madeira, S. Demeshko, G. Eckold, and J.
Jimenez-Gallegos, Z. Phys. Chem., \textbf{223,} 701-711(2009).

\bibitem {survey}E. Reguera, H. Yee-Madeira, J. Fern\'{a}ndez-Bertran, L.
Nu\v{n}ez, Transition Met. Chem., \textbf{24}, 163-167(1999).

\bibitem {Nomura}W. Kosaka, K. Nomura, K. Hashimoto, and S.-I. Ohkoshi, J. Am.
Chem. Soc. \textbf{127}, 8590 (2005).

\bibitem {Keggin}J. F. Keggin, and F.D. Miles, Nature (London) \textbf{137}, 577-578(1936).

\bibitem {Ludi}A. Ludi, H.U. G\"{u}del, Struct. Bonding, \textbf{14} 1-21, (1973).

\bibitem {Buser}H.J. Buser, D. Schwarzenbach, W. Petter, and A. Ludi, Inorg.
Chem. \textbf{16}, 2704-2710(1977).

\bibitem {Tokoro}H. Tokoro, S.-I. Ohkoshi, T. Matsuda and K. Hashimoto, Inorg.
Chem. \textbf{43}, 5231-5236 ( 2004).

\bibitem {RWZN}A.I. Rykov, J. Wang, T. Zhang and K. Nomura, Hyperfine
Interact. DOI 10.1007/s10751-012-0705-5 (in press).

\bibitem {Nozik}A.J. Nozik and M. Kaplan, Phys. Rev. \textbf{159}, 273 (1067).

\bibitem {Evans}R.J. Evans, D.G. Rancourt, and M. Grodzicki, American
Mineralogist \textbf{90}, 187-198 (2005).

\bibitem {Konig1}E. K\"{o}nig and A.S. Chakravarty, Theoret. Chim. Acta
(Berl.) \textbf{9}, 151--170 (1967).

\bibitem {Nature}B.N. Figgis, J. Lewis, F. Mabbs, and G.A. Webb, Nature
\textbf{203}, 1138 (1964).

\bibitem {Konig2}E. K\"{o}nig, A.S. Chakravarty, and K. Madeja, Theoret. Chim.
Acta (Berl.) \textbf{9}, 171--181 (1967).

\bibitem {Figgis}B.N. Figgis, J. Lewis, F. Mabbs, and G.A. Webb, J. Chem. Soc.
(A) 442-447 (1964).

\bibitem {Note}The results tabulated in the above works\cite{Konig2,Figgis}
are reported with different conventions for the sign of crystal field
splitting $\delta$. We avoid using the negative numbers for the crystal-field
splitting energy. Instead of $\left\vert \delta\right\vert $, and sign of
$\delta$ we use explicitly the notations of $\Delta_{1}$ and $\Delta_{2}$,
such that $\Delta_{1}=$ $\Delta_{2}$ when orbital singlet is lowest, and
$\Delta_{2}=0$, when the orbital singlet is lowest.

\bibitem {Ganiel}U. Ganiel, Chem. Phys. Lett., 4, 87 (1969).

\bibitem {Ingalls}R. Ingalls, Phys. Rev. \textbf{133,} A787-A795 (1964).

\bibitem {Merrithew}P.B. Merrithew, P.G. Rasmussen, and D.H. Vincent, Inorg.
Chem. \textbf{10}, 1401-1406 (1971).

\bibitem {Matsuda}T. Matsuda, J.E. Kim, K. Ohoyama, and Y. Moritomo, Phys.Rev.
\textbf{B 79}, 172302 (2009).

\bibitem {Mikuli}E. Mikuli, A. Migda\l -Mikuli, and J. Mayer, J. Thermal Anal.
\textbf{54}, 93-102(1998).

\bibitem {neutron1}C. N\"{o}ldeke, B. Asmussen, W. Press, H. B\"{u}ttner, G.
Kearley, R. E. Lechner and B. Ruffl\'{e}, J. Chem. Phys. \textbf{113} (2000) 3219.

\bibitem {sus}B.K. Chaudhuri, Solid St. Comm, \textbf{16}, 767-772 (1975).

\bibitem {Adak}S. Adak, L.L. Daemen, M. Hartl, D. Williams, J. Summerhill, H.
Nakotte, J. Solid St. Chem. \textbf{184}, 2854 (2011).

\bibitem {Reiff}M.W. Reiff, R.B. Frankel, B.F. Little, and G.J. Long, Chem.
Phys. Lett. \textbf{28, }68 (1974).

\bibitem {Widman}A. Widmann, H. Kahlert, I. Petrovic-Prelevic, H. Wulff, J. V.
Yakhmi, N. Bagkar, and F. Scholz, Inorg. Chem. \textbf{ 41} (2002) 5706-5715.

\bibitem {Ayrault}S. Ayrault, B. Jimenez, E. Garnier, M. Fedoroff, D.J. Jones,
and C. Loos-Neskovic, J. Solid St. Chem. \textbf{141}, 475-485 (1998).

\bibitem {Loos}C. Loos-Neskovic, S. Ayrault, V. Badillo, B. Jimenez, E.
Garnier, M. Fedoroff, D.J. Jones, and B. Merinov, J. Solid St. Chem.
\textbf{177}, 1817-1828 (2004).

\bibitem {Herren}F. Herren, P. Fischer, A. Ludi, and W. H\"{a}lg, Inorg. Chem.
\textbf{19}, 956-959 (1980) .

\bibitem {Kaye}S.S. Kaye, J.R. Long, Catal. Today, \textbf{120,} 311-316 (2007).

\bibitem {Gutlich}V. Ksenofontov, A. B. Gaspar and P G\"{u}tlich, Topics in
Current Chemistry, 2004, Volume 235, Spin Crossover in Transition Metal
Compounds III, Pages 39-66.

\bibitem {Wojdel}J.C. Wojde\l , I.P.R. Moreira, and F. Illas, J. Chem. Phys.
\textbf{130,} 014702 (2009).

\bibitem {Middlemiss}D.S. Middlemiss, D. Portinari, C.P. Grey, C.A. Morrison,
\ and C.C. Wilson, Phys. Rev. \textbf{B 81}, 184410 (2010).

\bibitem {Hayami}S. Hayami, Y. Shigeyoshi, M. Akita, K. Inoue, K. Kato, K.
Osaka, M. Takata, R. Kawajiri, T. Mitani, and Y. Maeda, Angew. Chem. Int. Ed.
\textbf{44}, 4899 (2005).

\bibitem {Manoharan}P.T. Manoharan, B. Sambandam, R. Amasarani, B. Varghese,
C.S. Gopinath, and K. Nomura, Inorg. Chem. Acta \textbf{374}, 586 (2011).

\bibitem {Petra}P.J. van Koningsbruggen, Y. Maeda, H. Oshio, Top. Curr. Chem.
\textbf{233}, 259 (2004).
\end{thebibliography}
\end{document}